\documentclass[prd,%
preprint,%
tightenlines,superscriptaddress,
nofootinbib,eqsecnum,amsfonts,amsmath,amssymb,bm]{revtex4}
\usepackage{bm}
\usepackage{graphicx}

\begin{document}

\title{Quantum effects in gravitational wave signals from cuspy superstrings}

\author{Diego Chialva} 
\affiliation{International School for Advanced Studies (SISSA), Via Beirut 2-4, I-34013 Trieste, Italy}
\affiliation{INFN Sezione di Trieste}

\author{Thibault Damour} 
\affiliation{Institut des Hautes \'Etudes Scientifiques, 35 route de Chartres, F-91440
Bures-sur-Yvette, France}

\begin{abstract}
We study the gravitational emission, in Superstring Theory, from fundamental strings exhibiting cusps. The {\it classical} computation of the gravitational radiation signal from cuspy strings features strong bursts in the special null directions associated to the cusps. We perform a {\it quantum} computation of the gravitational radiation signal from a cuspy string, as measured in a gravitational wave detector using matched filtering and located in the special null direction associated to the cusp. We study the quantum statistics (expectation value and variance)
 of the measured filtered signal and find that it is very sharply peaked around the classical prediction. Ultimately, this result follows from the fact that the detector is a low-pass filter which is blind to the violent high-frequency quantum fluctuations of both the string worldsheet $X^{\mu} (\tau , \sigma)$, and the incoming gravitational field $h_{\mu\nu}^{in} (x)$.
\end{abstract}

\date{\today}

\maketitle

\section{Introduction}\label{secI}


\smallskip

Recently, there has been a renewed interest in the possibility \cite{Witten85} that fundamental strings of Superstring Theory may have astronomical sizes and play the role of cosmic strings \cite{book}. There are indeed viable models of brane inflation \cite{DvaliTye,KKLMMT} where stable strings, of horizon size, are produced at the end of inflation \cite{Sarangi,Jones,Dvali,Copeland,Polchinski,Firouzjahi,Vilenkin}. The tension $\mu$ of these strings is expected \cite{Sarangi}, \cite {KKLMMT}, \cite{Copeland,Polchinski,Firouzjahi} to lie in the range $10^{-11} \lesssim G\mu \lesssim 10^{-6}$. The upper bound of this range is however already very constrained both by pulsar timing observations \cite{KTR94,MZVL96,Lommen} and by measurements of the anisotropy of the cosmic microwave background \cite{Pogosian}.

\smallskip

Until recently, it was thought that the gravitational effects of strings with tension $G \mu \ll 10^{-6}$ were too weak to be observable. However, it has been shown in \cite{DV1,DV2,Damour2} that strings with tensions in the large range $10^{-13} \lesssim G\mu \lesssim 10^{-6}$ (which includes the range expected from brane inflation models) could be detected by the gravitational wave interferometers LIGO and LISA through the observation of the {\it gravitational wave bursts} associated to cusp formation. It has long be known that cusps periodically form during the oscillatory evolution of a generic smooth string loop \cite{Turok}. Geometrically, a cusp corresponds to an (isolated) point on the string world-sheet where the tangent plane to the world-sheet is null (i.e. tangent to the light-cone) instead of being time-like, as it generically is. In other words, a future-directed local light-cone (with vertex on the world-sheet) generically intersects the world-sheet along two distinct null vectors. A cusp is a special point on the world-sheet where these two null vectors coincide (or are parallel). The common null direction, say $\ell^{\mu}$, of these coinciding null vectors defines the null direction of strongest emission of the gravitational wave bursts studied in \cite{DV1,DV2,Damour2}.

\smallskip

The crucial feature that makes the gravitational wave bursts associated to cusps sensitive probes of tensions as small as $G\mu \gtrsim 10^{-13}$ is the fact that, in the Fourier domain, their gravitational wave amplitude $h$ is proportional to the inverse cubic root of the frequency $f$ of observation: $h(f) \propto \vert f \vert^{-1/3}$ \cite{DV1,DV2}\footnote{This corresponds to a time-domain waveform proportional to $h(t) \propto \vert t-t_c \vert^{1/3}$. Note that $h(f)$ denotes here the {\it logarithmic} Fourier transform of $h(t) : h(f) \equiv \vert f \vert \, \tilde h (2 \pi f)$ where $\tilde h (\omega) \equiv \int dt \, e^{i \omega t} \, h(t)$ is the usual Fourier transform.}.

\smallskip

This (weak) power-law dependence on the frequency, $\propto \vert f \vert^{-1/3}$, is directly related to the (weak) geometrical singularity which exists at the isolated points of the world-sheet where cusps form. As the geometrical definition of these cusp singularities depends on a {\it classical} description of the world-sheet, one might worry that they be blurred by {\it quantum} fluctuation effects (associated to the effective string length\footnote{It will sometimes be convenient to consider units where $c=1$ but where $\hbar$ is not set to one, and to correspondingly define the string tension $\mu$ with (classical) units [mass]/[length]. In these units the combination $G\mu$ is dimensionless, independently of $\hbar$, and naturally enters classical gravitational wave calculations.} $\ell_s = (\hbar / (2 \pi \mu))^{1/2}$)\footnote{Let us note that Ref.~\cite{SO03} has shown that the waveform of gravitational wave bursts from cusps is robust against the presence of (classical) small-scale wiggles on the string.}. The main purpose of the present paper is to perform a {\it quantum} computation of the observable gravitational radiation signal from a cuspy string to investigate to what extent quantum effects might blur the special feature that makes the corresponding classical signal such a sensitive probe of small tensions. Our conclusion will be that quantum effects (both in the string dynamics which sources the signal, and in the emitted gravitational field) are utterly negligible and jeopardize in no way the measurability estimates made in \cite{DV1,DV2,Damour2}.

\smallskip

The computations of the present work are quite different from the ones
of previous studies of quantum effects in gravitational radiation from
superstrings \cite{CIR, Iengo}. Indeed previous studies considered the quantum spectrum of massless emission, i.e. the probability for certain massive string energy eigenstates to decay into another massive string energy eigenstate and a (unique) massless (graviton) state. Such computations would be mostly relevant if one had initially prepared the string into a particular energy eigenstate and had a detector that could observe individual outgoing gravitons. However, we are concerned here with quite a different physical situation. The detectors we are interested in (LIGO, LISA,$\ldots$) do not detect individual gravitons but measure instead a certain (quasi-classical) filtered wave amplitude. In addition, we shall argue that the massive string states we are interested in  are not typical energy eigenstates but, instead, some quasi-classical coherent states. As a consequence of this special physical situation we shall not be able to express our quantum computation within the usual string perturbation formalism, but will resort to a mixture of first-quantized strings and second-quantized gravitational field. We leave to future work a derivation of our approximate results from a fully consistent string theory framework.

\section{The two types of quantum effects in filtered gravitational wave signals from strings}\label{sec2}

Let us motivate our discussion by considering, as model problem, a second quantized field theory\footnote{The fact that superstring theory is, essentially, only defined as a {\it first} quantized theory of string states makes it technically difficult to start from string transition amplitudes to discuss all the quantum effects we want to discuss. This is why we find convenient to use such a second quantized field theory model.} where a massless field $h(x)$ is coupled to a {\it quantum} source $J(x)$,
\begin{equation}
\label{eq1n}
S = \int d^D x \left[ \frac{1}{2} \, h(x) \, \Box \, h(x) + h(x) \, J(x) + \cdots \right] \, ,
\end{equation}
where the ellipsis concern the dynamics of the quantum variables entering the definition of the source $J(x)$. [In our application the latter variables will be the string world-sheet coordinates $X^{\mu} (\tau , \sigma)$.] For simplicity, we suppress all Lorentz indices and write equations as if the field $h(x)$ were a scalar. We have, however, in mind a massless spin-$2$ field $h_{\mu\nu} (x)$ (so that the kinetic operator\footnote{We use the signature ``mostly plus''.} $\Box$ in equation (\ref{eq1n}) should be replaced by a suitably gauge-fixed version of the Einstein-Pauli-Fierz kinetic operator; we shall define our normalization of $h_{\mu\nu}$ below). The Heisenberg quantum equation of motion for the field $h(x)$ reads
\begin{equation}
\label{eq2n}
- \, \Box \, h(x) = J(x) \, .
\end{equation}
If $G_{\rm ret}$ denotes the retarded Green's function $(- \, \Box \, G_{\rm ret} (x) = \delta^D (x))$ the second quantized field operator $h(x)$ can be written in the form (see, e.g., \cite{IZ}\footnote{Chapter IV of the textbook \cite{IZ} considers a quantum field $h(x)$ interacting with a {\it classical} source $J(x)$. We simply extends here the use of the general (Heisenberg-picture) result (\ref{eq3n}) to the case of a {\it quantum} source $J(x)$, associated to the dynamics of a quantized string. Rigorously speaking, one would need to start from a second quantized approach to string theory to make full sense of the formulas we write. Their physical meaning (discussed below) is, however, so transparent that we are confident of the correctness (to leading order in $G\mu$) of our final results.})
\begin{equation}
\label{eq3n}
h(x) = h_{in} (x) + \int d^D \, y \, G_{\rm ret} (x-y) \, J(y) \, .
\end{equation}
In this equation the {\it in} field operator $h_{in} (x)$ is a free field $(\Box \, h_{in} (x) = 0)$ which describes the incoming vacuum fluctuations of $h(x)$. We work here in the Heisenberg picture and assume that the quantum state of the field $h(x)$ is the {\it in vacuum} $\vert 0 \rangle_{in}$.

\smallskip

In the following we shall consider a gravitational wave detector such as LIGO or LISA, and express its measurement in terms of the field $h(x)$, Eq.~(\ref{eq3n}), considered in the far radiation zone of the source $J$.

\smallskip

As always when discussing quantum effects, it is crucial to make clear what experimental situation, and precise observable, one is considering. Here, we are considering, as basic quantum observable, the {\it filtered output} of a gravitational wave detector looking for bursts of the form predicted by the (classical) computation of \cite{DV1,DV2}. The instantaneous output of a gravitational wave detector is (after suitable normalization) of the form
\begin{equation}
\label{eq5n}
o(t) = \zeta^{\mu\nu} \, \hat h_{\mu\nu} (t,\bm{x}_0) + n(t) \, ,
\end{equation}
where $n(t)$ is the noise in the detector, $\hat h_{\mu\nu} (x)$ a gauge-invariant projection\footnote{A gravitational wave detector is sensitive only to a gauge-invariant measure of the gravitational field. For instance, one can define $\hat h_{\mu\nu}$, in the rest frame of the detector, by $\hat h_{00} = 0 = \hat h_{0i}$ and $R_{0i0j} = -\frac{1}{2} \, \partial_{00} \, \hat h_{ij}$.} of the gravitational field of Eq.~(\ref{eq3n}), considered in the radiation zone, $\bm{x}_0$ the spatial location of the (center of mass of the) detector, and $\zeta^{\mu\nu}$ the suitably normalized\footnote{Say $\zeta^{\mu\nu} \, \zeta_{\mu\nu}^* = 1$.} polarization tensor to which the detector is sensitive. When looking, in the noisy time series of the output $o(t)$, for a signal having a certain classically predicted behaviour $h^{\rm predict} (t)$, $h(t) \propto \vert t-t_c \vert^{1/3}$, or, in the Fourier domain\footnote{We use here the usual Fourier transform $\tilde h (\omega) \equiv \int dt \, e^{i \omega t} \, h(t)$.} $\tilde h^{\rm predict} (\omega) \propto \vert \omega \vert^{-4/3}$, the optimal linear filter consists in considering as basic observable the {\it filtered output}
\begin{equation}
\label{eq6n}
o_f \equiv \int dt \, f(t) \, o(t) = \int dt \, f(t) \, \zeta^{\mu\nu} \, \hat h_{\mu\nu} (t,\bm{x}_0) + \int dt \, f(t) \, n(t)
\end{equation}
where $f(t)$ is a time-domain {\it filter function}\footnote{More precisely one considers a bank of time-translated filter functions $f(t-t_0)$, with varying ``arrival times'' $t_0$.} defined, in the Fourier domain, as
\begin{equation}
\label{eq7n}
\tilde f (\omega) = \frac{\tilde h^{\rm predict} (\omega)}{S_n (\omega)} \, ,
\end{equation}
where $\tilde h^{\rm predict} (\omega)$ is the Fourier transform of the predicted signal, and $S_n (\omega)$ the noise spectral density, i.e. the Fourier transform of the correlation function of the detector's noise:
\begin{equation}
\label{eq8n}
\langle n(t_1) \, n(t_2) \rangle = C_n (t_1 - t_2) = \int \frac{d\omega}{2\pi} \, S_n (\omega) \, e^{-i\omega (t_1 - t_2)} \, .
\end{equation}

When discussing the detection of gravitational wave bursts of the form predicted in \cite{DV1,DV2}, i.e. $h^{\rm predict} (t) \propto \vert t-t_c \vert^{1/3}$, one should define a filter function $f(t)$ by taking $\tilde h^{\rm predict} (\omega) = \vert \omega \vert^{-4/3}$ in Eq.~(\ref{eq7n}), and then taking the inverse Fourier transform: $f(t) = \int \frac{d\omega}{2\pi} \, e^{-i\omega t} \, \tilde f(\omega) = \int \frac{d\omega}{2\pi} \, e^{-i\omega t} \, \tilde h^{\rm predict} (\omega) / S_n (\omega)$. Note that the optimal filter $f(t)$ differs from the expected time-domain signal $h^{\rm predict} (t)$ by the convolution action of a ``whitening kernel'' with Fourier transform $1/S_n (\omega)$.

\smallskip

When inserting the decomposition (\ref{eq3n}) of the radiation field $h_{\mu\nu} (x)$ entering the filtered output of Eq.~(\ref{eq6n}), we see that the basic quantum observable $o_f$ can be decomposed into three contributions:
\begin{equation}
\label{eq9n}
o_f = A_f^{in} + A_f^J + n_f \, ,
\end{equation}
where, after introducing the $J$-generated field entering Eq.~(\ref{eq3n}) namely,
\begin{equation}
\label{eq10n}
h_{\mu\nu}^J (x) \equiv \int d^D \, y \, G_{\rm ret} (x-y) \, J_{\mu\nu} (y) \, ,
\end{equation}
we have defined
\begin{equation}
\label{eq11n}
A_f^{in} \equiv \int dt \, f(t) \, \zeta^{\mu\nu} \, \hat h_{\mu\nu}^{in} (t,\bm{x}_0) \, ,
\end{equation}
\begin{equation}
\label{eq12n}
A_f^J \equiv \int dt \, f(t) \, \zeta^{\mu\nu} \, \hat h_{\mu\nu}^J (t,\bm{x}_0) \, ,
\end{equation}
\begin{equation}
\label{eq13n}
n_f \equiv \int dt \, f(t) \, n(t) \, .
\end{equation}
Viewed in the vacuum $\vert 0 \rangle_{in}$ appropriate to the Heisenberg picture we are using, the in field $h_{\mu\nu}^{in}$ has vanishing expectation value. Therefore, the first contribution $A_f^{in}$, Eq.~(\ref{eq11n}), which we might call the ``filtered in amplitude'', describes the effect of the quantum vacuum fluctuations in $h(x)$, as seen in the filtered output. In other words, the fluctuating quantum observable $A_f^{in}$ describes the shot noise due to the quantized nature of the gravitational field $h_{\mu\nu} (x)$\footnote{Our use of a second quantized description of the gravitational field $h_{\mu\nu} (x)$ allowed us to derive with ease the expression of the contribution $A_f^{in}$ to the observed filtered signal $o_f$. A derivation of $A_f^{in}$ within first-quantized string perturbation theory would be more cumbersome because one would need to study multi-graviton amplitudes.}. The second contribution $A_f^J$, Eq.~(\ref{eq12n}), which is the ``filtered gravitational wave amplitude generated by the source $J$'', will be the main focus of this work, in the case where the source $J_{\mu\nu}$ is the stress-energy tensor $T_{\mu\nu}$ of a cosmic-size superstring. Finally, the third contribution $n_f$, Eq.~(\ref{eq13n}), is simply the filtered effect of the detector's noise (which is usually treated as a classical random variable).

\smallskip

Note that, contrary to the two contributions $A_f^{in}$ and $n_f$ which are purely fluctuating, i.e. have zero expectation values, the contribution $A_f^J$ generated by the quantum stress-energy tensor $J_{\mu\nu} \propto T_{\mu\nu} (X)$ of a superstring (considered in a highly excited state) will have a non-zero expectation value (hopefully corresponding to the classical computation of \cite{DV1,DV2}) around which quantum effects linked to the quantized dynamics of the string will introduce fluctuations. The aim of our computation is to quantitatively estimate the magnitude of the latter fluctuations, i.e. the probability distribution function of the quantum observable $A_f^J$, to see to what extent quantum fluctuations of the string dynamics might blur the classically expected signal.

\section{Light-cone gauge description of cuspy string states in superstring theory}\label{sec3}

In order to compute the probability distribution function of the filtered, source-generated gravitational wave amplitude $A_f^J$, Eq.~(\ref{eq12n}) with Eq.~(\ref{eq10n}), we need two ingredients: (i) a description of the quantum operator $J_{\mu\nu}$ in terms of the string dynamical variables, and (ii) a description of the quantum state of a cuspy string. In this section, we shall discuss this second ingredient.

\smallskip

Let us first remark that the experimental situation we are interested in is a special one. We consider a state where, as seen from the small subsystem we are interested in (made of a massive string and of some gravitational excitations\footnote{i.e. some extra massless string states.}), a macroscopic external agency (the cosmological expansion) has stretched an initially microscopic-size string state into a quasi-classical, macroscopic-size state. If the system we were interested in was (to give a simple example) an harmonic oscillator $\hat X (t)$, we would describe this blowing up, by a time-dependent external agency, to the macroscopic level, of an initially microscopic quantum state (say the ground state), by coupling the harmonic oscillator to a large, classical, time-dependent external force $F(t)$. It is then well-known that the final state of the oscillator is described (in the Schr\"odinger picture) by a {\it coherent state}, namely: $\vert F \rangle = \exp i \int dt \, F(t) \, \hat X (t) \, \vert 0 \rangle$, where $F(t)$ is the classical force and $\hat X (t) = (2 \, \omega_0)^{-1/2} (\hat a \, e^{-i\omega_0 t} + \hat a^{\dagger} (t) \, e^{i \omega_0 t})$ the position operator of the (unit mass) oscillator. Thinking of the string as a collection of oscillators, this simple example motivates us to considering that an appropriate description of the quantum state of a superstring stretched to macroscopic sizes by the cosmological expansion is a certain {\it coherent state}\footnote{We shall work henceforth within the approximation where, when discussing the distribution function of the observable $A_f^J$, Eq.~(\ref{eq12n}), the coupling of the string state to gravity (and, in particular, its decay under back reaction) is neglected, so that we consider that the state of the string is some given coherent state which determines the statistics of $A_f^J$. This is similar to saying that, in ordinary Schr\"odinger quantum mechanics, the statistics of some observable $f(X)$ is determined by the wave function $\psi (X;t)$, supposedly known at the time $t$ where the observable $f(X)$ is observed.} of the infinite set of oscillators describing $X^{\mu} (\tau , \sigma)$. However, we need to take care of the constraints which gauge-away the ``time-like'' and ``longitudinal'' oscillators to leave only the $D-2$ transverse oscillators. The simplest way to do so is to work in a light-cone gauge
\begin{equation}
\label{eq3.1n}
n_{\mu} \, X^{\mu} (\tau , \sigma) = \alpha' (n_{\mu} \, p^{\mu}) \, \tau \, ,
\end{equation}
where $n_{\mu}$ is a certain fixed null vector. In this gauge, if we choose, say,  $n_{\mu} = \left( \frac{1}{\sqrt 2} , 0 , \ldots , 0 , \frac{1}{\sqrt 2} \right)$ so that $n_{\mu} \, X^{\mu} \equiv (X^0 + X^{D-1}) / \sqrt 2 \equiv X^+$, the $X^+$ oscillators are set to zero $(X^+ = \alpha' \, p^+ \, \tau)$, the $X^i$ ``transverse'' oscillators (with $i = 1,2,\ldots D-2$) are unconstrained, and the $X^-$ oscillators are expressed as quadratic combinations of the infinite set of transverse oscillators.

\smallskip

We shall therefore consider string states of the $\vert \alpha \rangle_R \, \vert \tilde\alpha \rangle_L$ where
\begin{eqnarray}
\label{eq3.2n}
\vert \alpha \rangle_R &= &\prod_{\{n\}, \{i\}} e^{-\frac{|\alpha^i_n|^2}{2} +\alpha^i_n a^{i \dagger}_n}|0\rangle \\
\vert \alpha \rangle_L &= &\prod_{\{n\}, \{i\}} e^{-\frac{|\tilde\alpha^i_n|^2}{2} +\tilde\alpha^i_n \tilde a^{i \dagger}_n}|0\rangle \, . \nonumber
\end{eqnarray}
Beware of the somewhat unconventional notation used in this equation. The quantities $\alpha_n^i$ (with $i=1,2,\ldots , D-2$ and $n = 1,2,\ldots$) denote some given $c$-numbers parametrizing the right-moving part of the considered coherent state, while the $a_n^i$, $a_n^{i\dagger} \equiv a_{-n}^i$ denote the annihilation and creation operators of the $n^{\rm th}$ right-moving mode. They are such that $[a_n^i , a_m^{j\dagger}] = [a_n^i , a_{-m}^j] = \delta^{ij} \, \delta_{n-m}^0$. The quantities $\tilde\alpha_n^i$, $\tilde a_n^i$ denote the corresponding quantities for the left-moving part of the string. In other words, the derivatives of the transverse string coordinates read
\begin{equation}
\label{eq3.3n}
 \partial_{\pm} \, X^i(\tau\pm\sigma)=\sqrt{\frac{\alpha'}{2}}\sum_{n}\sqrt{n}
 \ (\!\!~^(\tilde a^{)i}_n \, e^{-in(\tau\pm\sigma)}+ \!\!~^({}\tilde a^{)i}_{-n} \, e^{in(\tau\pm\sigma)})
\end{equation}
where $\sigma^{\pm} \equiv \tau \pm \sigma$, $\partial_{\pm} \equiv \partial / \partial \sigma^{\pm} = \frac{1}{2} \, (\partial_{\tau} \pm \partial_{\sigma})$, and where the `tilde' quantities correspond to $\partial_+$ and $\tau + \sigma$ (left-movers).

\smallskip

Because of the Lorentz covariance of light-cone quantization (in the critical dimension\footnote{In the formal developments we try to keep a general $D$ and assume $D=10$ (for the superstring). However, we shall later assume that the considered coherent state has a macroscopic extension only in $D=4$ uncompactified dimensions. The compactified dimensions will correspond to small values of the $\alpha$'s and $\tilde\alpha$'s and will have a negligible contribution to our final results.}) we have full freedom in selecting the special null direction $n^{\mu}$ entering Eq.~(\ref{eq3.1n}). We shall follow here the following logic. We start from some {\it given} coherent state, i.e. two sequences of complex numbers $\alpha_n^i , \tilde\alpha_n^i$. These numbers define {\it both} a {\it quantum} state {\it and} a {\it classical} solution of the string equations of motion (in a generic light-cone gauge (\ref{eq3.1n})). We know from \cite{Turok} that, generically, this classical solution will exhibit cusps, with certain associated null directions $\ell^{\mu} (\alpha , \tilde\alpha)$ of intense classical gravitational radiation emission. To simplify our computation of the quantum observable $A_f^J$ associated to some cusp null direction $\ell^{\mu} (\alpha,\tilde\alpha)$, we wish to ensure that $\ell^{\mu} (\alpha,\tilde\alpha)$ is parallel to $n^{\mu}$\footnote{Actually, three different special null vectors will enter our study: the cusp null direction $\ell^{\mu}$, the light-cone gauge null direction $n^{\mu}$ and the ``graviton momentum'' $k_{\omega}^{\mu}$ (see below). We recall that, in the light-cone gauge, the operator $X^- (\tau , \sigma) = (X^0 - X^{D-1}) / \sqrt 2$ is a complicated (quadratic) functional of the transverse oscillators. To avoid having a term $\exp (i \, k^+ \, X^-)$ in the source-generated $h_{\mu\nu}^J$ (see below) we need $0 = k^+ = k^{\mu} \, n_{\mu}$, which implies the parallelism of the two real null vectors $k^{\mu}$ and $n^{\mu}$. Finally, as we wish to study gravitational emission in the ``cusp'' direction $\ell^{\mu}$, we shall have to require that the three null vectors $\ell^{\mu}$, $n^{\mu}$ and $k^{\mu}$ are all parallel.}. Such a parallelism can always be realized (simply by rotating appropriately $n^{\mu}$). However, after such a rotation, the values of the sequences of $c$-numbers $\alpha_n^i , \tilde\alpha_n^i$ will change in a complicated manner. Actually, the new ``specially aligned'' sequences $(\alpha^{\rm new} , \tilde\alpha^{\rm new})$ must now behave in a non generic way, that we need to specify.

\smallskip

We recall (see, e.g., \cite{DV2}) that, in a generic conformal gauge $\sigma_+^{\rm old} , \sigma_-^{\rm old}$, the left and right string coordinates\footnote{We follow here \cite{DV2} in writing
$X^{\mu} = \frac{1}{2}( X_{+}^{\mu} + X_{-}^{\mu}).$}
behave, near a cusp (located at $X^{\mu} = 0$ and occurring at $\sigma_+ = 0 = \sigma_-$), as
\begin{equation}
\label{eq3.4n}
X_{\pm}^{\mu} (\sigma_{\pm}^{\rm old}) = \ell^{\mu} \, \sigma_{\pm}^{\rm old} + \frac{1}{2} \, \ddot X_{\pm}^{\mu} (\sigma_{\pm}^{\rm old})^2 + \frac{1}{6} \, X_{\pm}^{(3)\mu} (\sigma_{\pm}^{\rm old})^3 + \cdots
\end{equation}
so that (remembering that $\ell_{\mu} \, \ddot X_{\pm}^{\mu} = 0$ at the cusp)
\begin{equation}
\label{eq3.5n}
\ell_{\mu} \, X_{\pm}^{\mu} (\sigma_{\pm}^{\rm old}) = \frac{1}{6} \, \ell_{\mu} \, X_{\pm}^{(3)\mu} (\sigma_{\pm}^{\rm old})^3 + \cdots
\end{equation}
On the other hand, in the specially aligned ``new'' light-cone gauge $n^{\mu} \propto \ell^{\mu}$, we must have, from Eq.~(\ref{eq3.1n}), the property that 
\begin{equation}
\label{eq3.6n}
\ell_{\mu} \, X_{\pm}^{\mu} =  \alpha' (\ell_{\mu} \, p^{\mu}) \, \sigma_{\pm}^{\rm new} \, .
\end{equation}
The conclusion is therefore that $\sigma_{\pm}^{\rm new} \sim (\sigma_{\pm}^{\rm old})^3$. Considering the transverse components of $X_{\pm}^{\mu}$, Eq.~(\ref{eq3.4n}), in the new gauge, we then have
\begin{equation}
\label{eq3.7n}
X_{\pm}^i (\sigma_{\pm}^{\rm new}) = \frac{1}{2} \, \ddot X_{\pm}^i (\sigma_{\pm}^{\rm old})^2 + \cdots \sim (\sigma_{\pm}^{\rm new})^{2/3} \, ,
\end{equation}
so that
\begin{equation}
\label{eq3.8n}
\partial_{\pm} \, X_{\pm}^i (\sigma_{\pm}^{\rm new}) \sim (\sigma_{\pm}^{\rm new})^{-1/3} \, .
\end{equation}
Note that, geometrically, it means that the transverse projection of the string $X^i (\tau^{\rm new} , \sigma^{\rm new})$ draws, at the light-cone moment where the cusp forms $(\tau^{\rm new} = 0)$, a cuspy curve in transverse space. This is a simple geometrical consequence of having aligned our light-cone gauge precisely with the null direction associated with the cusp.

\smallskip

The result (\ref{eq3.8n}) applies to the classical expectation value of $\partial_{\pm} \, X_{\pm}^i$. For a coherent state, this is simply given by replacing the operators $a_n^i , \tilde a_n^i$ in Eq.~(\ref{eq3.3n}) by $\alpha_n^i , \tilde\alpha_n^i$ respectively. This means that the sequences of $c$-numbers $\alpha_n^i , \tilde\alpha_n^i$ must have a certain power-law behaviour
\begin{equation}
\label{eq3.9n}
\alpha_n^i \sim \tilde \alpha_n^i \sim n^{-\gamma}
\end{equation}
as $n \to \infty$ with (considering the right-moving modes)
\begin{equation}
\label{eq3.10n}
\partial_- \, X^i (\sigma_-) \propto \sum_n \sqrt n \, \alpha_n^i \, e^{-in\sigma_-} \sim \sum_{n=1}^{1/\sigma_-} n^{\frac{1}{2} - \gamma} \sim \left( \frac{1}{\sigma_-} \right)^{\frac{3}{2} - \gamma} \, .
\end{equation}
Comparing to Eq.~(\ref{eq3.8n}) determines the power index $\gamma$ of (\ref{eq3.9n}) to be
\begin{equation}
\label{eq3.11n}
\gamma = \frac{7}{6} \, .
\end{equation}
In conclusion, we shall consider coherent string states of the form (\ref{eq3.2n}) with sequences of $c$-numbers $\alpha_n^i , \tilde\alpha_n^i$ satisfying (\ref{eq3.9n}) with (\ref{eq3.11n}). Such states will describe a quantum version of a cusp aligned with our chosen direction of observation $n^{\mu} \propto \ell^{\mu}$.

\section{Gravitational wave signals from cuspy string states}\label{sec4}

We turn now to a description of the filtered string-generated signal $A_f^J$, Eq.~(\ref{eq12n}), with $h_{\mu\nu}^J$ given by Eq.~(\ref{eq10n}). Before discussing the exact form of the source $J_{\mu\nu}$ corresponding to the gravitational coupling of the string, let us consider the effect of the retarded Green's function $G_{\rm ret}$ in the integral (\ref{eq10n}). When considering a string state which is localized around some center of mass worldline $y^{\mu} \simeq x_0^{\mu} (\tau)$ (at a cosmological distance from the Earth), and a detector located in the solar system, we can approximate the (Fourier transformed) retarded Green's function in Eq.~(\ref{eq10n}) (considered for concreteness in the physically relevant case of $D=4$ uncompactified dimensions)
\begin{equation}
\label{eq4.1n}
G_{\rm ret} (t_x , \bm{x} ; t_y , \bm{y}) = \int \frac{d\omega}{2\pi} \, G_{\omega}^{\rm ret} (\bm{x} ; \bm{y}) \, e^{-i\omega (t_x - t_y)} \, ,
\end{equation}
in the following well-known way
\begin{equation}
\label{eq4.2n}
G^{\rm ret}_{\omega} (\bm{x} ; \bm{y}) = \frac{e^{i\omega \vert \bm{x} - \bm{y} \vert}}{4\pi \vert \bm{x} - \bm{y} \vert} \simeq \frac{e^{i\omega \vert \bm{x} \vert}}{4\pi \vert \bm{x} \vert} \, e^{-i\omega \bm{n} \cdot \bm{y}} \, ,
\end{equation}
where $\bm{n} \equiv \bm{x} / \vert \bm{x} \vert$ is the unit vector directed from the origin (located, say, at the center of mass of the string) towards the detector.

\smallskip

Inserting Eqs.~(\ref{eq4.1n}), (\ref{eq4.2n}) into Eq.~(\ref{eq10n}) yields (see \cite{DV2})
\begin{equation}
\label{eq4.3n}
h_{\mu\nu}^J (t,\bm{x}) \simeq \frac{1}{4\pi r} \int \frac{d\omega}{2\pi} \, e^{-i\omega (t-r)} \tilde J_{\mu\nu} (\omega , \omega \bm{n})
\end{equation}
where $r \equiv \vert \bm{x} \vert$ and where $\tilde J_{\mu\nu} (k^{\lambda})$, with $k^{\lambda} = (\omega , \bm{k})$, is the spacetime Fourier transform of the source $J_{\mu\nu} (x)$:
\begin{equation}
\label{eq4.4n}
\tilde J_{\mu\nu} (k^{\lambda}) = \int d^D \, x \, e^{-i k_{\lambda} x^{\lambda}} J_{\mu\nu} (x) \, .
\end{equation}
Note the important fact that the integral in (\ref{eq4.3n}) contains only an integration over the frequency $\omega$. The integral over $d^3 \bm{k}$ has been effectively already performed and has led to the fact, apparent on the right-hand side (rhs) of Eq.~(\ref{eq4.3n}), that the only $\bm{k}$'s entering the final result are related to the frequency by $\bm{k} = \omega \bm{n}$. In other words $k^{\mu} = (\omega , \bm{k}) = (\omega , \omega \bm{n})$ is a null vector directed from the source towards the detector.

\smallskip

If we normalize the gravitational field $h_{\mu\nu} (x)$ in the geometrical (einsteinian) way, i.e. $g_{\mu\nu} (x) = \eta_{\mu\nu} + h_{\mu\nu} (x)$, the source $J_{\mu\nu}$ of $h_{\mu\nu}$ (with $-\Box \, h_{\mu\nu} = J_{\mu\nu}$) will be 
\begin{equation}
\label{eq4.5n}
J_{\mu\nu} (x) = +16 \, \pi \, G \left( T_{\mu\nu} (x) - \frac{1}{D-2} \, \eta_{\mu\nu} \, T (x) \right)
\end{equation}
where $G$ is Newton's constant, and where $T_{\mu\nu} (x)$ denotes the stress-energy tensor of the source. Then, we can write $h_{\mu\nu}^J$ as (in $D=4$)
\begin{equation}
\label{eq4.6n}
h_{\mu\nu}^J (t,\bm{x}) \simeq \frac{4 \, G}{r} \int \frac{d\omega}{2\pi} \, e^{-i\omega (t-r)} \left( \tilde T_{\mu\nu} (k_{\omega}^{\lambda}) - \frac{1}{2} \, \eta_{\mu\nu} \, \tilde T (k_{\omega}^{\lambda}) \right) \, ,
\end{equation}
where $\tilde T_{\mu\nu} (k^{\lambda})$ is the spacetime Fourier transform of $T_{\mu\nu} (x^{\lambda})$, and where $k_{\omega}^{\lambda} = (\omega , \omega \bm{n})$.

\smallskip

The coupling $\int \frac{1}{2} \, h_{\mu\nu} \, T^{\mu\nu}$ between a {\it fundamental}\footnote{The effects linked to the quantum fluctuations around the Nambu-Goto dynamics investigated here are clearly present (possibly together with other effects) in all types of strings: fundamental, Dirichlet or even gauge-theory ones.} string, of tension $\mu$\footnote{The tension $\mu$ denotes the effective four-dimensional tension, including eventual warp factors.} and the gravitational field is proportional to
\begin{eqnarray}
\label{eq4.6nbis}
\mu \int d^2\sigma &&\bigl[ h_{\mu\nu}(X)(\partial_+ X^\mu\partial_-X^\nu)+
 h_{\mu\nu}(X)(\psi_+^\mu\partial_-\psi_+^\nu)+
 h_{\mu\nu}(X)(\psi_-^\mu\partial_+\psi_-^\nu) \nonumber \\
 &&+  \ R_{\mu\nu\rho\sigma}(X)\psi_+^\mu\psi_+^\nu\psi_-^\rho\psi_-^\sigma\bigl] \, .
 \end{eqnarray}
 Because of the on-shell constraint $\psi_{\pm} = \psi_{\pm} (\tau \pm \sigma)$ the terms $\psi^{\mu}_+ \, \partial_- \, \psi_+^{\nu}$ and $\psi^{\mu}_- \, \partial_+ \, \psi_-^{\nu}$ do not contribute to $T^{\mu\nu}$. As for the term quartic in the fermions, it does, a priori, contribute a term $\propto (\delta \, R_{\alpha\beta\gamma\delta} / \delta \, h_{\mu\nu}) \, \psi_+^{\alpha} \, \psi_+^{\beta} \, \psi_-^{\gamma} \, \psi_-^{\delta}$. This term contributes to the Fourier-transformed $\tilde T_{\mu\nu} (k^{\lambda})$ a term proportional to $(k \cdot \psi_+) (k \cdot \psi_-) \, \psi_+^{\mu} \, \psi_-^{\nu}$.
 
 \smallskip
 
 We shall further simplify our computation by considering a detector which is, as seen from the source, precisely at the center of the gravitational burst emitted by the cusp. In other words, we shall require that the basic null direction $n^{\mu}$ defining the light-cone gauge (Eq.~(\ref{eq3.1n})) is not only parallel to the null direction $\ell^{\mu}$ defined by the cusp, but also to the null direction $(1,\bm{n})$ connecting the source to the detector. This implies that all the graviton momenta $k_{\omega}^{\mu} = (\omega , \omega \bm{n})$ entering the radiation field (\ref{eq4.6n}) are also parallel to $n^{\mu}$. As one sets $\psi_{\pm}^+ = 0$ in the light-cone gauge, $k \cdot \psi_{\pm} = -k^- \psi_{\pm}^+$ vanishes in the light-cone gauge.
 
\smallskip

Finally, it is enough to consider the bosonic contribution to the stress-energy tensor $(d^2 \sigma \equiv d\tau \, d\sigma)$
\begin{eqnarray}
\label{eq4.7n}
T^{\mu\nu} (x) &= &\mu \int d^2 \sigma \, \delta^D (x-X(\tau,\sigma)) (\partial_{\tau} \, X^{\mu} \, \partial_{\tau} \, X^{\nu} - \partial_{\sigma} \, X^{\mu} \, \partial_{\sigma} \, X^{\nu}) \nonumber \\
&= &4\mu \int d^2 \sigma \, \delta^D (x-X(\sigma_+ , \sigma_-)) \, \partial_+ \, X^{(\mu} \, \partial_- \, X^{\nu)} \, ,
\end{eqnarray}
and to its Fourier transform
\begin{equation}
\label{eq4.8n}
\tilde T^{\mu\nu} (k) = 4\mu \int d^2 \sigma \, e^{-ik\cdot X} \, \partial_+ \, X^{(\mu} \, \partial_- \, X^{\nu)} \, ,
\end{equation}
which is the vertex operator for the emission or absorption of a graviton of momentum $k$ by a string.

\smallskip

If we simplify formulae by re-writing the filter function $f(t)$ in Eq.~(\ref{eq12n}) as $f^{\rm new} (t-r_0)$ (where $r_0 = \vert \bm{x}_0 \vert$ is the radial distance from the source to the detector), the filtered source-generated amplitude (\ref{eq12n}) reads (taking into account the tracelessness of $\zeta^{\mu\nu}$)
\begin{equation}
\label{eq4.9n}
A_f^J = \frac{4G}{r_0} \int \frac{d\omega}{2\pi} \, \tilde f (-\omega) \, \zeta_{\mu\nu} \, \hat{\tilde T}^{\mu\nu} (k_{\omega}^{\lambda}) \, ,
\end{equation}
where $\tilde f (\omega) = \int dt \, e^{+ i \omega t} f^{\rm new} (t)$ is the Fourier transform of $f^{\rm new} (t) = f^{\rm old} (t+r_0)$. As explained above the ``hat'' over $\tilde T^{\mu\nu}$ denotes the projection over the gauge-invariant gravitational wave amplitude affecting the detector (which is only sensitive to tidal forces). We can use the remaining freedom in our light-cone frame to ensure that the detector is ``at rest'' with respect to the usual Lorentz frame $(x^0 , x^1 , x^2 , x^3)$ behind the light-cone frame (i.e. with $x^+ = (x^0 + x^3) / \sqrt 2$). Then the detector's polarization tensor will have only spatial components $\zeta_{IJ}$, $I,J = 1,2,3$. In addition, the ``hat'' projection simply consists in projecting $h_{\mu\nu}$ on its physically active {\it transverse traceless} (TT) components $h_{ij}$, $i,j = 1,2$ only. Let us define, as usual, the TT projection operator of a symmetric spatial tensor $k_{IJ}$, $I,J = 1,2,3$ by $k_{IJ}^{\rm TT} \equiv \left( P_{IK} \, P_{JL} - \frac{1}{2} \, P_{IJ} \, P_{KL} \right) k_{KL}$ with $P_{IJ} \equiv \delta_{IJ} - n^I \, n^J$, where $n^I = \bm{n}$ is, as above, the unit direction vector from the source towards the detector. One easily checks that only $\zeta_{IJ}^{\rm TT}$ matters in Eq.~(\ref{eq4.9n}), and that its only non zero components are transverse: $\zeta_{ij}^{\rm TT}$, with $i,j=1,2$ only. [Note that we have aligned our frame so that $n^I = \delta_3^I$ points in the third, longitudinal direction.] We can then rewrite the result (\ref{eq4.9n}) purely in terms of transverse components\footnote{To avoid any confusion: beware that the transverse components of $\zeta_{ij}^{\rm TT}$ differ from the restriction $\zeta_{ij}$ of $\zeta_{IJ}$ to its transverse components $i,j = 1,2$.} $i,j=1,2$
\begin{equation}
\label{eq4.10n}
A_f^J = \frac{4G}{r_0} \int \frac{d\omega}{2\pi} \, \tilde f (-\omega) \, \zeta_{ij}^{\rm TT} \, \tilde T^{ij} (k_{\omega}^{\lambda}) \, .
\end{equation}

The phase factor $e^{-ik\cdot X}$ entering the vertex operator (\ref{eq4.8n}) reads simply (in view of the definition (\ref{eq3.1n}) of the light-cone gauge and of the fact that our light-cone gauge ``null vector'' $n^{\mu}$ is aligned with $k^{\mu}$) $e^{-i\alpha' (k \cdot p)\tau} = e^{+i\alpha' M \omega \tau}$, where we used $k = (\omega , \omega \bm{n})$, $p = (M,\bm{0})$, where $M$ is the total mass energy of the string. Using also $\mu = 1/(2\pi \alpha')$, i.e. $\alpha' = 1/(2\pi \mu)$, the phase factor reads $e^{i \ell \omega \tau / (2\pi)}$ where we defined the invariant length of the string: $\ell \equiv M/\mu$.

\smallskip

Inserting the vertex operator (\ref{eq4.8n}) into (\ref{eq4.10n}) then leads to the following explicit expression for our basic filtered gravitational wave signal
\begin{equation}
\label{eq4.11n}
A_f^J = \frac{16 G\mu}{r_0} \int d\tau \, d\sigma \, \frac{d\omega}{2\pi} \, \tilde f (-\omega) \, \zeta_{ij}^{\rm TT} \, e^{i\ell \omega \tau / (2\pi)} \, \partial_+ \, X^i \, \partial_- \, X^j \, ,
\end{equation}
in which one should insert the oscillator expansions (\ref{eq3.3n})\footnote{Note that, contrary to what happened in the time-gauge calculation of Refs. \cite{DV1}, \cite{DV2}, there is not any more a ``cubic'' saddle point in the phase factor $e^{-ik\cdot X}$ of Eq.~(\ref{eq4.11n}). The fact that a cusp is a strong emitter of high-frequency gravitational waves shows up, when using a ``cusp-aligned'' light-cone gauge, in the singular behaviour $\partial_{\pm} \, X \sim (\sigma_{\pm})^{-1/3}$ of the string coordinate gradients entering the vertex operator $\partial_+ \, X^i \, \partial_- \, X^j$.}. 

\smallskip

Eqs.~(\ref{eq4.11n}) and (\ref{eq3.3n}) define the quantum observable $A_f^J$ as a certain operator in the string dynamics Hilbert space. It is easy to perform explicitly the triple integration in Eq.~(\ref{eq4.11n}): (i) the integral over $\sigma$ yields a Kronecker $\delta_{nm}$ between the left and right mode numbers $e^{-in(\tau + \sigma)}$, $e^{-im(\tau - \sigma)}$; (ii) the integral over $\tau$ then yields some delta functions of the frequency $\delta \left( \frac{\ell \omega}{2\pi} - 2n \right)$ or $\delta \left( \frac{\ell \omega}{2\pi} + 2n \right)$; and (iii) the integral over $\omega$ then yields a series over $n$. Finally, defining (as in \cite{DV2}) the fundamental frequency $\omega_1 \equiv 4\pi / \ell$, we get
\begin{equation}
\label{eq4.12n}
A_f^J = 16 \pi \, \frac{G}{\ell r_0} \, \zeta_{ij}^{\rm TT} \sum_{n=1}^{\infty} n \, [\tilde f (-n \, \omega_1) \, a_n^i \, \tilde a_n^j + \tilde f (n \, \omega_1) \, a_n^{i\dagger} \, \tilde a_n^{j\dagger}] \, .
\end{equation}

\section{Quantum noise contributions to gravitational wave signals}\label{sec5}

Eq.~(\ref{eq9n}) exhibited a decomposition of the filtered output, $o_f$, of a gravitational wave detector into three terms: (i) the quantum noise $A_f^{in}$, Eq.~(\ref{eq11n}), which describes the filtered vacuum fluctuations of the gravitational field, (ii) the filtered, string-generated signal $A_f^J$, explicitly expressed as Eq.~(\ref{eq4.12n}), and (iii) the filtered detector noise $n_f$, Eq.~(\ref{eq13n}). Let us consider, for concreteness, the case of the LIGO detector. As discussed in \cite{DV2}, and recalled above, the optimal filter for detecting gravitational wave bursts from cuspy strings is Eq.~(\ref{eq7n}) with $\tilde h^{\rm predict} (\omega) \propto \vert \omega \vert^{-4/3} \cdot e^{i \omega t_0}$. The division by $S_n (\omega)$ provides a Fourier-domain filter $\tilde f (\omega)$, Eq.~(\ref{eq7n}), which is peaked around the characteristic (circular) frequency $\omega_* \simeq (2\pi) \times 150$ Hz \cite{DV2}. It is then convenient to normalize the filter function $f(t)$ entering Eqs.~(\ref{eq11n}), (\ref{eq12n}), (\ref{eq13n}) so that the modulus of $\tilde f (\omega_*) = \int dt \, f(t) \, e^{i\omega_* t}$ is equal to one. With this normalization, Eq.~(6.5) of \cite{DV2} says that the filtered detector noise $n_f$ of initial LIGO can be roughly modelled as a Gaussian variable with standard deviation $\sigma_n^f \simeq 1.7 \times 10^{-22}$. Advanced LIGO might reach a level smaller by a factor $13.5$, i.e. $\sigma_n^f \simeq 1.3 \times 10^{-23}$. Fig.~1 of Ref.~\cite{DV2} (and their multi-parameter generalizations in Ref.~\cite{Damour2}) shows that the classical estimate of the string-generated signal $A_f^J$ might be comparable (and hopefully larger) than these noise levels in a wide range of string tensions $10^{-12} \lesssim G\mu \lesssim 10^{-6}$. Let us now estimate the quantum statistics of, both, $A_f^{in}$ and $A_f^J$, to see whether quantum noise can play any significant role.

\smallskip

Let us start by considering the vacuum fluctuation term $A_f^{in}$, Eq.~(\ref{eq11n}), which reads, when using a TT gauge ($I,J=1,2,3$),
\begin{equation}
\label{eq5.1n}
A_f^{in} = \int dt \, f(t) \, \zeta^{IJ} \, h_{IJ}^{in{\rm TT}} (t, \bm{x}_0) \, .
\end{equation}
Here, $h_{IJ}^{in{\rm TT}} (x)$ is a free gravitational radiation field, considered in its vacuum state. Therefore, $A_f^{in}$, Eq.~(\ref{eq5.1n}), is a quantum Gaussian noise. To compute its standard deviation, let us use the two-point (Wightman) correlation function of $h^{in} (x)$:
\begin{equation}
\label{eq5.2n}
\langle h_{IJ}^{in{\rm TT}} (t,\bm{x}) \, h_{I'J'}^{in{\rm TT}} (t' , \bm{x}') \rangle = 32 \, \pi \, G \int \frac{d^3 k}{(2\pi)^3 \, 2 \, \omega_k} \, P_{IJI'J'}^{\rm TT} (\hat{\bm{k}}) \, e^{-i\omega_k (t-t') + i\bm{k} (\bm{x} - \bm{x}')}
\end{equation}
where the factor $32 \, \pi \, G$ comes from the ``geometric'' (instead of ``canonical'') normalization of $h_{\mu\nu}$ and where $P^{\rm TT}$ denotes the TT projector used above (here considered for the direction $\hat{\bm{k}} = \bm{k} / \omega_k$ with $\omega_k = \vert \bm{k} \vert$). We then find that the variance of $A_f^{in}$ (using $\zeta^{IJ} \, \zeta_{IJ} = 1$) is of order
\begin{equation}
\label{eq5.3n}
(\sigma (A_f^{in}))^2 \sim G \int d\omega \, \omega \vert \tilde f (\omega) \vert^2 \sim G \, \omega_*^2 \vert \tilde f (\omega_*) \vert^2 \, ,
\end{equation}
where we use the fact that the Fourier filter $\tilde f (\omega)$ is peaked around $\omega_*$. As said above, we normalize the filter $f(t)$ so that $\vert \tilde f (\omega_*) \vert = 1$. Hence, the standard deviation of the filtered gravitational vacuum noise is of order
\begin{equation}
\label{eq5.4n}
\sigma (A_f^{in}) \sim G^{1/2} \, \omega_* = 2\pi \, f_* \, t_{\rm Planck} \sim 5 \times 10^{-41} \left( \frac{f_*}{150 \, {\rm Hz}} \right) \, .
\end{equation}

This is clearly too small to worry about. 

\smallskip

Let us now consider the statistical properties of the string-generated signal (\ref{eq4.12n}). As explained above, we model the state of the string by a coherent state $\vert \alpha \rangle_R \, \vert \tilde\alpha \rangle_L$. In such a state the expectation values of the annihilation and destruction operators $a_n^i , \tilde a_n^i , a_n^{i\dagger} , \tilde a_n^{i\dagger}$ are, by definition, the $c$-numbers $\alpha_n^i , \tilde\alpha_n^i$, and their complex conjugates (c.c.) $\bar\alpha_n^i$, $\bar{\tilde\alpha}_n^i$, so that the expectation value of $A_f^J$ reads
\begin{equation}
\label{eq5.5n}
\langle A_f^J \rangle = 16 \pi \, \frac{G}{\ell r_0} \, \zeta_{ij}^{\rm TT} \sum_{n=1}^{\infty} (n \, \tilde f (-n \, \omega_1) \, \alpha_n^i \, \tilde\alpha_n^j + {\rm c.c.}) \, .
\end{equation}

This result is evidently equivalent to replacing the operator $X^i$ in Eq.~(\ref{eq4.11n}) by its classical value (obtained by replacing $a$'s by $\alpha$'s in Eqs.~(\ref{eq3.3n})), and therefore equivalent to the results of Refs.~\cite{DV1,DV2}.

\smallskip

The principal novel result of this study is now obtained by considering the {\it variance} of the operator $A_f^J$ in the coherent state $\vert \alpha \rangle \, \vert \tilde\alpha \rangle$. Using $[a_n^i , a_m^{j\dagger}] = [\tilde a_n^i , \tilde a_n^{j\dagger}] = \delta^{ij} \, \delta_{nm}$ (and $[a,\tilde a] = 0$ etc.), it is easily computed from Eq.~(\ref{eq4.12n}) and found to be
\begin{equation}
\label{eq5.6n}
(\sigma (A_f^J))^2 =  \left(16 \pi \frac{G}{\ell r_0} \right)^2 \sum_{n=1}^{\infty} n^2 \vert \tilde f (n \, \omega_1)\vert^2 \, [\zeta_{is}^{\rm TT} \, \zeta_{js}^{\rm TT} (\alpha_n^i \, \bar\alpha_n^j + \tilde\alpha_n^i \bar{\tilde\alpha}_n^j) +  \zeta_{ij}^{\rm TT} \, \zeta_{ij}^{\rm TT}] \, .
\end{equation}

\smallskip

The first important thing to notice in the variance (\ref{eq5.6n}) is its crucial dependence on the detector's filter function $f(t)$. If one had considered a time-sharp filter, $f(t) = \delta (t-t_0)$, i.e. a frequency-flat $\vert \tilde f (\omega) \vert = 1$, the variance (\ref{eq5.6n})  would be {\it infinite}. Both the last, state-independent ``vacuum'' contribution $(\propto \! \Sigma \, n^2)$, and the first, state-dependent one $(\propto \Sigma \, n^2 (\vert \alpha_n \vert^2 + \vert \tilde\alpha_n \vert^2))$ would diverge for a cuspy string state (i.e. $\vert \alpha_n \vert^2 \sim \vert \tilde\alpha_n \vert^2 \sim n^{-7/3}$ according to Eqs.~(\ref{eq3.9n}), (\ref{eq3.11n})). Note, by contrast, that the expectation value (\ref{eq5.5n}) would remain convergent (like $\Sigma \, n^{-4/3}$) even for a time-sharp filter function. This shows that quantum fluctuations in a gravitational wave ``cusp'' burst are pretty violent on short time scales. On the other hand, if one takes into account the fact that the detecting filter is well peaked\footnote{The detector's noise spectrum $S_n (\omega)$ increases like $\omega^2$ for large frequencies. Therefore the filter function (\ref{eq7n}) decreases like $\omega^{-10/3}$.} around some optimal frequency $\omega_*$, we see from Eq.~(\ref{eq5.6n}) that the variance of the filtered cusp signal will be {\it finite} and of order (we use $\vert f(\omega_*)\vert = 1$, $(\zeta^{\rm TT})^2 \sim 1$ and consider $\Delta n \sim n_*$ terms around the peak value $n_*$ such that $n_* \, \omega_1 = \omega_*$)
\begin{equation}
\label{eq5.8n}
(\sigma (A_f^J))^2 \sim \left( 16 \pi \frac{G}{\ell r_0} \right)^2 n_*^3 (\vert \alpha_{n_*} \vert^2 + \vert \tilde\alpha_{n_*} \vert^2 + 1) \, .
\end{equation}
Comparing this result with the expectation value (\ref{eq5.5n}), i.e. $\langle A_f^J \rangle \sim (16 \pi G/(\ell r_0)) \, n_*^2 \vert \alpha_{n_*} \vert \, \vert \tilde\alpha_{n_*} \vert$, we can approximately write the standard deviation of $A_f^J$ in the form
\begin{equation}
\label{eq5.9n}
\sigma (A_f^J) \sim \frac{\vert\langle A_f^J \rangle \vert}{n_*^{1/2}} \left( \frac{1}{\vert \alpha_{n_*} \vert^2} + \frac{1}{\vert \tilde\alpha_{n_*} \vert^2} + \frac{1}{\vert \alpha_{n_*} \vert^2 \vert \tilde\alpha_{n_*} \vert^2} \right)^{1/2} \sim \frac{\vert \langle A_f^J \rangle \vert}{n_*^{1/2} \vert \alpha_{n_*} \vert} \, .
\end{equation}
In the last expression we have assumed that $\vert \tilde\alpha_{n_*} \vert \sim \vert \alpha_{n_*} \vert \gg 1$, so that one can neglect the ``vacuum'' contribution.

\smallskip

Let us now estimate how large $n_*$ and $\vert \alpha_{n_*} \vert$ are for the type of cosmic superstring that one might expect to detect via their gravitational wave bursts. Correspondingly to the assumption $\vert \ddot X_{\pm}^{\mu} \vert^{\rm time \, gauge} \sim 2\pi / \ell$ of Ref.~\cite{DV2} (i.e. that the string is not too wiggly), we can assume that, in the special cusp-related light-cone gauge we are using, the asymptotic behaviour $\vert \alpha_n \vert \sim \vert \tilde\alpha_n \vert \sim A / n^{\gamma}$ with $\gamma = 7/6$ is roughly valid from $n=1$ to infinity\footnote{We recall that the stretching effect of the cosmological expansion, together with the smoothing effects of loop production and, possibly, radiation damping, are expected to lead to a network made of loops whose overall shape is dominated (in the cosmic time gauge) by the first few modes $\alpha_n , \tilde\alpha_n$. The fact that such string states might look rather special within the ensemble of possible quantum string states (see, e.g., \cite{Iengo}) is not necessarily relevant within the physical context that we consider.}. Then we can use the closed string mass formula
\begin{equation}
\label{eq5.10n}
\frac{1}{4} \, \alpha' \, M^2 = \sum n \, a_n^{i\dagger} \, a_n^i = \sum n \, \tilde a_n^{i\dagger} \, \tilde a_n^i \sim \sum_{n=1}^{\infty} \vert A \vert^2 \, n^{-4/3} \, ,
\end{equation}
together with $\alpha' = 1/(2\pi \, \mu)$ and $M = \mu \, \ell$ to estimate the coefficient $A$: $\vert A \vert^2 \sim \mu \, \ell^2$. Using also the link $\omega_* \sim 4\pi \, n_* / \ell$, we have the estimates $n_* \sim f_* \, \ell$ and $\vert \alpha_{n_*} \vert \sim \ell \sqrt\mu \, (f_* \, \ell)^{-7/6}$ where $f_* \equiv \omega_* / 2\pi$ denotes the detector's optimal frequency. In other words the {\it ratio} between the standard deviation and the expectation value of $A_f^J$ can be estimated as
\begin{equation}
\label{eq5.11n}
{\mathcal R} = \frac{\sigma (A_f^J)}{\vert \langle A_f^J \rangle \vert} \sim \frac{1}{n_*^{1/2} \vert \alpha_{n_*}\vert} \sim f_* \, \ell_s \, (f_* \, \ell)^{-\frac{1}{3}} \sim f_* \, t_{\rm Planck} \, (f_* \, \ell)^{-\frac{1}{3}} (G\mu)^{-\frac{1}{2}} \, ,
\end{equation}
where $\ell_s \sim \mu^{-1/2}$ is the quantum string length, and $\ell \equiv M/\mu$ the invariant length of the considered coherent-state macoscopic string.

\smallskip

In terms of cosmologically relevant dimensionless ratios\footnote{Note that recent work \cite{Vanchurin:2005pa} suggests that $\ell$ is only ten times smaller than the cosmological horizon.} this yields
\begin{equation}
\label{eq5.12n}
{\mathcal R} \sim 10^{-43} \left( \frac{\ell}{10^{10} \, yr} \right)^{-\frac{1}{3}} \left( \frac{f_*}{150 \, {\rm Hz}} \right)^{\frac{2}{3}} \left( \frac{G\mu}{10^{-9}} \right)^{-\frac{1}{2}} \, .
\end{equation}
As this is a ratio, the corresponding absolute value of $\sigma (A_f^J)$ (for a detectable amplitude $\langle A_f^J \rangle \sim 10^{-22}$) will be down to the $\sim 10^{-65}$ level ! This is clearly negligibly small, even with respect to the already negligibly small graviton-shot noise term (\ref{eq5.4n}).

\smallskip

In summary, the main conclusions of this study are: The gravitational radiation ``cusp'' signal $h^{\rm cusp} (f_*)$ emitted by a string (in a coherent state) and detected by a low-frequency detector (with characteristic frequency $f_*$) such as LIGO or LISA, is affected by two types of quantum noise\footnote{We consider only the quantum effects carried by the gravitational wave signal $h_{\mu\nu}$. There are evidently many (much more relevant) quantum noise effects coming from the measuring process in the detector.}. On the one hand, the graviton shot noise $\delta h^{\rm shot} (f_*) \sim f_* \, t_{\rm Planck}$ (where $t_{\rm Planck} = (\hbar G)^{1/2}$), and, on the other hand, the signal coming from quantum fluctuations of the string stress-energy tensor near a cusp $\delta h^{\rm string} (f_*) \sim G\mu (f_* \, \ell)^{\frac{1}{3}} \, \frac{\ell_s}{r_0}$, (where $\ell_s \sim (\hbar / \mu)^{1/2})$. Both types of noise (and especially the second one) are totally negligible compared to the classical signal $h^{\rm cusp} (f_*) \sim G\mu (f_* \, \ell)^{-1/3} \, \frac{\ell}{r_0}$. The negligible character of the quantum noises crucially depend on considering the ``frequency windowing'' due to the detector, around a characteristic frequency $f_*$. Indeed, if the detector were allowed to make time-sharp measurements, i.e. if we allow the bandwidth $\Delta f \sim f_*$ to go to infinity, both types of quantum noises would diverge like a positive power of $f_*$. In other words, our computation highlights the fact that the observable cusp signal does not come from short distance scales (UV) near the cusp singularity (which do undergo violent fluctuations), but from length scales of order $f_*^{-1}$ which are intermediate between the UV scales $\sim \ell_s$ and the IR cut-off $\sim \ell$ associated to the overall size of the string.

\end{document}